\documentclass{article}
\usepackage{amsfonts}
\usepackage{graphicx}
\usepackage{caption}
\usepackage{subcaption}
\usepackage{amsmath}
\usepackage{mathtools}
\usepackage{geometry}
\usepackage{amssymb}

\usepackage{biblatex}

\allowdisplaybreaks

\let\oldref\ref
\renewcommand{\ref}[1]{(\oldref{#1})}
\begin{document}

\begin{center}
\Huge Chimeras in the two-community Kuramoto model with an external drive 
\bigskip
\normalsize

Jens Gr\o nborg
\bigskip

\end{center}

\begin{abstract}
\noindent
We study the bifurcations of a special case of the Kuramoto model with two communities of oscillators and an external drive. 
We use Ott-Antonsen's ansatz to derive the low-dimensional system of differential equations that governs the macroscopic dynamics of the high-dimensional problem. 
The choice of parameters of the system is motivated by the search for so-called Chimera states;
stable phase configurations with partial synchronization \cite{kuramoto2002coexistence,PhysRevLett.93.174102}.
Our main result is the derivation of the low-dimensional system following Ott-Antonsens Ansatz 
and findings of periodic and chaotic Chimeras.
\end{abstract}
\section*{Introduction}
The Kuramoto model has since its introduction in 1984 \cite*{ku} found application in scientific fields with interacting oscillators dynamics.
Oscillators appear in a wide range of areas from modeling Josephson Junctions \cite*{PhysRevE.67.026216, PhysRevLett.76.404} used in constructing superconducting integrated circuit chips for quantum computing, 
to describing modern power grid configurations \cite*{buttner2023complex}. In ecology the Kuramoto model is used to model the interaction between pests found on coffee plants \cite*{doi:10.1098/rsos.210122} 
and in neuroscience where neuron dynamics are characterized by chemical oscillators \cite*{10.3389/fnhum.2010.00190, odor,10.3389/fnetp.2022.910862}.
\\\\
Ott and Antonsens 2008 paper \cite{oa} marked a breakthrough in the study of Kuramoto-like models with a large number of oscillators.
The paper discusses the original model as well as variations on it. They all have global dynamics governed by a low-dimensional system under a series of constraints on the initial configuration of the system.
This paper is motivated by combining two of these variations \cite[p.9-11]{oa}; the community generalization with an external drive
\begin{equation}\label{ku}
d\theta_{\sigma k}/dt =
\omega_{\sigma k} +
\sum_{\sigma' = 1}^2
\frac{K_{\sigma\sigma'}}{N_{\sigma'}}
\sum_{j = 1}^{N_{\sigma'}}
\sin(\theta_{\sigma'j} - \theta_{\sigma k} - \beta) + \Lambda\sin(\Omega t- \theta_{\sigma k}).
\end{equation}
$\theta_{\sigma k}$ is the phase of an oscillator $k$ belonging to the community $\sigma$. $\omega_{\sigma k}$ is its initial frequency, 
$K_{\sigma\sigma'}$ the coupling constants between communities $\sigma$ and $\sigma'$ and  $\beta$ the phase distortion factor. 
The external drive is characterized by a weight $\Lambda$ and a frequency $\Omega$. 
The phase distortion parameter $\beta$ acts as a destabilizer by moving the fixed points of the coupling function away from the synchronized state enabling more complex dynamics \cite*{sa}.
\section*{Ott-Antonsen Ansatz}
The continuum case of Kuramoto-like models where $N\rightarrow\infty$ has received much attention in parts due to its complexity and in part its applicability.
Ott and Antonsen's paper \cite*{oa} sets off by considering the oscillators from the point of view of statistical mechanics.
The starting point is the Vlasov equation which is a conservation model from Plasma Physics. It models the conservation of oscillators and discounts collision effects, the same conditions that underpin the Kuramoto model 
\begin{equation}\label{5}
\partial f_\sigma/ \partial t + \partial / \partial \theta (f_\sigma \cdot v_\sigma) = 0.
\end{equation}
The phase and frequency of each oscillator are modeled as a continuous random variable with density function $f_\sigma(\omega_\sigma, \theta_\sigma, t)$.
The velocity $v_\sigma$ term is given by the Kuramoto model.
Ott and Antonsen proceeds by defining a so-called order parameter $r = 1/N\sum e^{i\theta}$ is introduced which is used to rewrite \ref{ku} as a single equation
\begin{equation}\label{kured}
d\theta_\sigma/dt =
\omega_\sigma - \Omega +
\frac{1}{2i}\sum_{\sigma' = 1}^2
(K_{\sigma\sigma'} r_{\sigma'}e^{-i\beta} + \Lambda) e^{-i\theta_\sigma} - c.c.,
\end{equation}
where c.c. is the complex conjugate of the prior term.
$r$ encodes the synchronicity of a community of oscillators in its absolute value and is in the continuum case defined in terms of density function $f$ 
\begin{equation}\label{r}
r_\sigma = \int_{-\infty}^\infty \int_0^{2\pi} f_\sigma e^{i\theta_\sigma} \mathrm{d}\theta_\sigma\mathrm{d}\omega_\sigma.
\end{equation}
Following the steps from Ott and Antonsen \cite{oa} we expand the density $f_\sigma$ as a Fourier series in the phase variable 
\begin{equation}\label{fu}
f_\sigma = \frac{g_\sigma(\omega)}{2\pi}(1+\sum_{n=1}^\infty f_{\sigma n}(\omega_\sigma, t)e^{in\theta_\sigma} + c.c.),
\end{equation}
Ott-Antonsen chose the Fourier coefficients as a power series of a real function $\alpha_\sigma$
\begin{equation}
f_{\sigma n}(\omega_\sigma,t) = (\alpha_\sigma(\omega_\sigma,t))^n.
\end{equation}
With this special choice of coefficients, we insert \ref{kured} and \ref{fu} in \ref{5} 
\begin{multline}\label{1d}
\partial f_\sigma/ \partial t + \partial / \partial \theta (f_\sigma \cdot v_\sigma) = 
\partial\alpha_\sigma/ \partial t \sum_{n=1}^{\infty}n\alpha_\sigma^{n-1}e^{in\theta_\sigma} + c.c. \\
+
(\omega_\sigma-\Omega)\sum_{n=1}^{\infty}in\alpha_\sigma^ne^{in\theta_\sigma} + c.c. \\
+ 
\left(\frac{1}{2i}\sum_{\sigma'=1}^2(K_{\sigma\sigma'} r_{\sigma'}e^{-i\beta} + \Lambda)e^{-i\theta_\sigma} - c.c. \right)
\sum_{n=1}^{\infty}in\alpha_\sigma^ne^{in\theta_\sigma} + c.c. \\
- 
\left(\frac{1}{2}\sum_{\sigma'=1}^2(K_{\sigma\sigma'} r_{\sigma'}e^{-i\beta} + \Lambda)e^{-i\theta_\sigma} + c.c. \right)
\left(1+\sum_{n=1}^\infty \alpha_\sigma^ne^{in\theta_\sigma} + c.c.\right) \\
=
\partial\alpha_\sigma/ \partial t \sum_{n=1}^{\infty}n\alpha_\sigma^{n-1}e^{in\theta_\sigma} + c.c. \\
+
(\omega_\sigma-\Omega)\sum_{n=1}^{\infty}in\alpha_\sigma^ne^{in\theta_\sigma} + c.c. \\
+
\frac{1}{2}\sum_{n=1}^{\infty}\sum_{\sigma'=1}^2(K_{\sigma\sigma'} r_{\sigma'}e^{-i\beta} + \Lambda)(n-1)\alpha_\sigma^ne^{i(n-1)\theta_\sigma} - \sum_{\sigma'=1}^2(K_{\sigma\sigma'} r_{\sigma'}e^{-i\beta} + \Lambda)(n+1){\alpha_\sigma^n}^*e^{-i(n+1)\theta_\sigma}  \\
- 
\sum_{\sigma'=1}^2(K_{\sigma\sigma'} r_{\sigma'}e^{-i\beta} + \Lambda)^*(n+1)\alpha_\sigma^ne^{i(n+1)\theta_\sigma} + \sum_{\sigma'=1}^2(K_{\sigma\sigma'} r_{\sigma'}e^{-i\beta} + \Lambda)^*(n-1) {\alpha_\sigma^n}^*e^{-i(n-1)\theta_\sigma}  \\
- 
\left(\frac{1}{2}\sum_{\sigma'=1}^2(K_{\sigma\sigma'} r_{\sigma'}e^{-i\beta} + \Lambda)e^{-i\theta_\sigma} + c.c. \right) \\
=
\partial\alpha_\sigma/ \partial t \sum_{n=1}^{\infty}n\alpha_\sigma^{n-1}e^{in\theta_\sigma} + c.c. 
+
(\omega_\sigma-\Omega)i\alpha_\sigma\sum_{n=1}^{\infty}n\alpha_\sigma^{n-1}e^{in\theta_\sigma} + c.c. \\
+
\left(\frac{1}{2}\sum_{\sigma'=1}^2\alpha_\sigma^2(K_{\sigma\sigma'} r_{\sigma'}e^{-i\beta} + \Lambda) -
(K_{\sigma\sigma'} r_{\sigma'}e^{-i\beta} + \Lambda)^*\right)\sum_{n=1}^{\infty}n\alpha_\sigma^{n-1}e^{in\theta_\sigma} + c.c. = 0
\end{multline}
We notice that surprisingly the $\theta$-dependency can be factored out leaving
\begin{equation}\label{1e}
\partial\alpha_\sigma/ \partial t + 
i(\omega_\sigma-\Omega)\alpha_\sigma + 
\frac{1}{2}\sum_{\sigma'=1}^2
(K_{\sigma\sigma'} r_{\sigma'}e^{-i\beta} + \Lambda) \alpha_\sigma^2 - (K_{\sigma\sigma'} r_{\sigma'}e^{-i\beta} + \Lambda)^* = 0.
\end{equation}
Despite the simplification, eq. \ref{r} and \ref{1e} which determines the dynamics of the global order parameter $r$ is still an infinite-dimensional problem because $\alpha$ is a function.
To address this choose the frequency density $g$ to be a Lorentzian, $g_\sigma(\omega) = \Delta_\sigma/\pi((\omega_\sigma-\omega_{\sigma0}^2)+\Delta_\sigma^2)^{-1}$,
and the integrals in \ref{r} have a closed analytic form
\begin{equation}
r_\sigma = \int_{-\infty}^\infty \int_0^{2\pi} f_\sigma e^{i\theta} \mathrm{d}\theta_\sigma\mathrm{d}\omega_\sigma= \int_{-\infty}^\infty \alpha(\omega,t)^*g(\omega) \mathrm{d}\omega_\sigma = \alpha_\sigma(\omega_{\sigma0}-i\Delta_\sigma,t)^*.
\end{equation}
With this result \ref{1e} simplifies to a set of complex ODEs
\begin{equation}\label{1f}
\partial r_\sigma/ \partial t = 
\Delta_\sigma+ i(\omega_{\sigma0}-\Omega)r_\sigma^* + 
\frac{1}{2}\sum_{\sigma'=1}^2
(K_{\sigma\sigma'} r_{\sigma'}^*e^{-i\beta} + \Lambda) r_\sigma^2 - (K_{\sigma\sigma'} r_{\sigma'}^*e^{-i\beta} + \Lambda)^*.
\end{equation}

\section*{Numerical analysis}
\begin{figure}
\centering
\begin{subfigure}{.5\textwidth}
\centering
  \includegraphics[width=\linewidth]{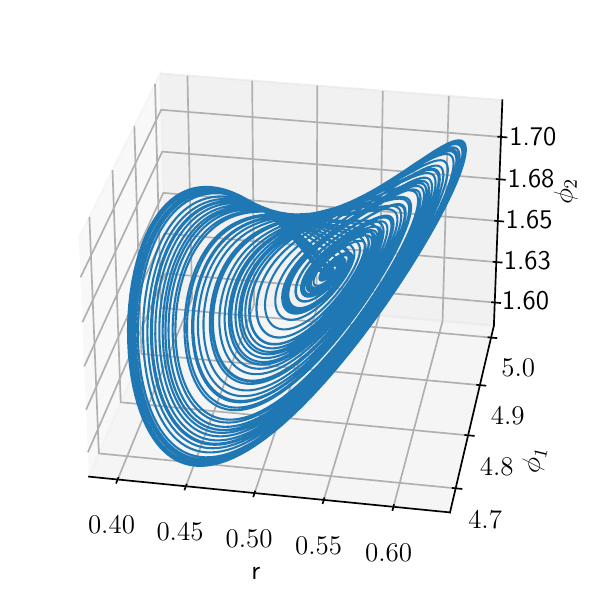}
  \caption{Shilnikov repeller after homoclinic bifurcation.}
  \label{fig:b}
\end{subfigure}%
\begin{subfigure}{.5\textwidth}
\centering
  \includegraphics[width=\linewidth]{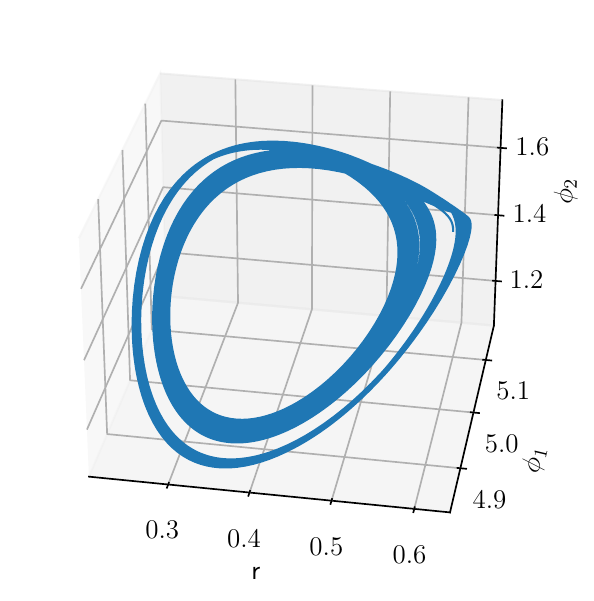}
  \caption{Stable chaotic attractor after a period-doubling cascade.}
  \label{fig:c}
\end{subfigure}%
\caption{Chaotic solutions.}
\end{figure}

Any numerical analysis of \ref{1f} will be local in nature and non-exhaustive. 
Our focus is to show that \ref{ku} has stable Chimera configurations in the continuum limit. 
We make parameter reductions in line with what can be found in related works \cite{PhysRevLett.93.174102, Bick_2018}.
\\\\
We simplify the search by choosing an initial value that is a Chimera. This is achieved by letting the first community of oscillators be synchronized,
$|r_1| = 1$, and let the oscillators be identical by pinching the scale parameter $\Delta_\sigma \rightarrow 0$. 
\\\\
We only consider the case when the internal and external coupling constants are symmetric
$K_{12}=K_{21}, K_{11} = K_{22}$. We normalize the two and define a coupling constant discrepancy parameter $A$, $K_{11} + K_{12} = 1, A \equiv K_{11}-K_{12}$.
It is convenient to shift the Sakaguchi-parameter $\beta$ such that $\beta \rightarrow \pi/2 - \beta$ and 
define a second discrepancy parameter between the frequencies of the external drive and
the oscillators initial frequency $\omega_\sigma \equiv \Omega - \omega_{0\sigma}$. 
These constraints reduces \ref{1f} to 
\begin{equation}\label{kuf}
\begin{aligned}
dr/dt &= \frac{1-r^2}{4}\left( (1-A)\sin(\phi_1-\phi_2+\beta) + (1+A) r\sin(\beta)+2\Lambda \cos(\phi_2) \right) \\
d\phi_1/dt &= \omega_1   - \frac{1-A}{2} r\cos(\phi_2 -\phi_1 + \beta)-\frac{1+A}{2}\cos(\beta) -\Lambda \sin(\phi_1)\\
d\phi_2/dt &=  \omega_2-\frac{1+r^2}{4r}\left((1-A)\sin(\phi_1 -\phi_2+\beta)+(1+A) r\cos(\beta)+2\Lambda \sin(\phi_2) \right).
\end{aligned}
\end{equation}
where $r = |r_2|$ in order to simplify the notation.
\begin{figure}
\centering
\includegraphics[width=\linewidth]{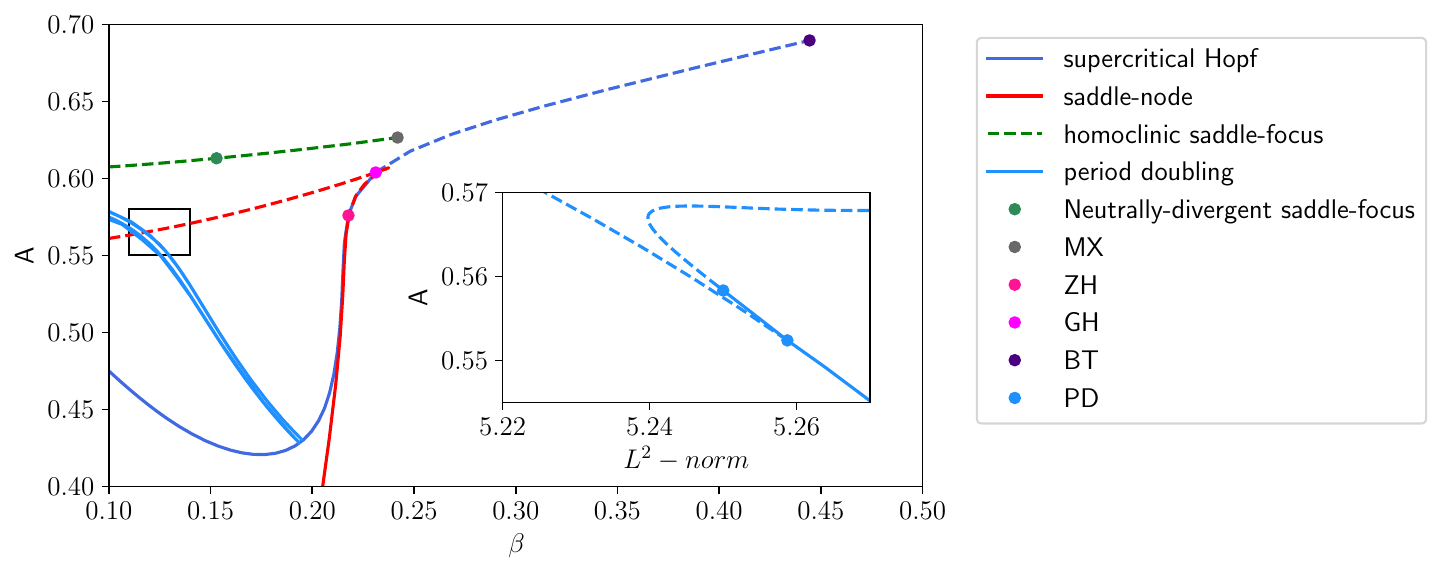}
\caption{Bifurcations in the $(A,\beta)$ space for fixed parameters $\omega_1 = 0.203, \omega_2 = 0.9, \Lambda = 0.5$, dotted lines mark unstable bifurcations.
The subdiagram shows how the original Hopf cycle loses stability in subsequent period doublings in the $(A,L^2-norm)$ space.}
\label{fig:a}
\end{figure}
\\\\
Solutions to this system exist in a torus-shaped subset of $\mathbb{R}^3$,
Chimera states are found in the interior of the torus, as solutions on the surface represent a completely synchronized system with $r = 1$.
Fixed points of \ref{kuf} can be found in the interior and are the basis for bifurcation diagrams as seen in Figure \ref{fig:a}. 
\\\\
We use the software packages auto-07p \cite*{Doedel07auto-07p:continuation} and xppaut \cite*{xpp} to continue bifurcations and investigate state space solutions. 
Figure \ref{fig:a} shows a complicated cut in the $(A,\beta)$ parameter space which we will discuss.
\\\\
Reading the diagram in a clockwise fashion a subcritical Hopf bifurcation originates from the Bogdanov-Takens (BT) point. 
When $\beta$ is decreased the Hopf line eventually intersects the saddle-node bifurcation in a Generalized-Hopf (GH) point and becomes supercritical.
The now supercritical Hopf line intersects the saddle-node a second time, this time tangentially in a Zero-Hopf (ZH) point. We also observe that this marks the beginning of a stable saddle-node bifurcation.
\\\\
A fixed point attractor can be found in the parameter space between the stable saddle-node line and the supercritical Hopf line.
This solution is a Chimera state, as the second community for this set of parameter values is in a stable desynchronized state.
The degree of desynchronization, as measured by $r$, is however constant. 
\\\\
When the discrepancy $A$ between the interior and exterior coupling constants is increased the fixed point attractor undergoes a Hopf bifurcation resulting in an oscillating Chimera state.
A further increase leads to a loss of stability in a period-doubling (PD) bifurcation. 
For a region of $\beta \in [0.1,0.14]$ we observe a cascade of period-doublings resulting in a stable chaotic attractor.
The magnified diagram in Figure \ref{fig:a} shows this stability transition for increasing $A$ and fixed $beta = 0.13$.
A Solution on the chaotic attractor which appears after the second period-doubling is shown in figure \ref{fig:c}. 
\\\\
We continue the original cycle that lost stability in the first period-doubling by increasing $A$ further, and eventually observe a rapidly increasing period indicating the presence of a homoclinic orbit. 
We use auto-07p's Homcont module to continue a high period cycle towards the Bogdonov-Takens point. Unfortunately, Homcont was not able to continue all the way leading to a convergence failure (MX) when the homoclinic closes in on the subcritical Hopf line.
\\\\
Homcont can be configured to detect test functions equivalent to definitions found in \cite*[p. 528]{Kuznetsov1998}. 
The saddle point to the green homoclinic line is of the type saddle-focus, and for $\beta \gtrapprox 0.15$ the eigenvalues $\lambda_1,\mu_1,\mu_2$ have values such that  
$\lambda_1 + \mathrm{Re}(\mu_1) + \mathrm{Re}(\mu_2) = 0$. 
To the left of this neutrally-divergent saddle-focus point, the homoclinic is a single orbit, and to the right, it is a chaotic repeller of the Shilnikov type. See Figure \ref{fig:b}.
\\\\
This concludes our numerical findings. We have shown how the Kuramoto model with communities and an external drive defined in \ref{ku} can be analyzed using Ott and Antonsens Ansatz. 
Using their method we derived the low-dimensional system, and analyzed it hoping to find Chimera states. We found that the model supports all the different types of Chimeras, including chaotic ones which came about from period-doubling cascade.
\newpage
\clearpage
\pagestyle{empty}

\addcontentsline{toc}{chapter}{References}

\renewcommand{\thepage}{}

\printbibliography

\end{document}